\pdfoutput=1
\documentclass[aps, twocolumn, floatfix, prb]{revtex4-1}
\usepackage{graphicx}
\DeclareGraphicsExtensions{.pdf}
\usepackage{amsmath,amssymb,bbold}
\usepackage{float}
\usepackage{bm}
\usepackage{afterpage}
\usepackage{color}
\usepackage{comment}
\usepackage{bm}
\usepackage[colorlinks=true,breaklinks=true,linkcolor=blue,anchorcolor=red,citecolor=blue,urlcolor=blue]{hyperref}

\newcommand{\bk}{{\bm k}}
\newcommand{\bU}{{\bm U}}
\newcommand{\bR}{{\bm R}}

\newcommand{\br}{{\bm r}}
\newcommand{\bB}{{\bf B}}

\newcommand{\bK}{{\bf K}}

\begin{document}

\title{Strain-induced pseudo-magnetic field in \texorpdfstring{$\alpha-{\cal T}_3$}{} lattice}

\author{Junsong Sun}
\affiliation{School of Physics, Beihang University,
Beijing, 100191, China}

\author{Tianyu Liu}
\affiliation{Institute for Quantum Science and Engineering and Department of Physics, Southern University of Science and Technology, Shenzhen 518055, China}

\author{Yi Du}
\affiliation{School of Physics, Beihang University,
Beijing, 100191, China }

\author{Huaiming Guo}
\email{hmguo@buaa.edu.cn}
\affiliation{ School of Physics, Beihang University,
Beijing, 100191, China}

\begin{abstract}
We investigate the effects of a nonuniform uniaxial strain and a triaxial strain on the $\alpha-{\cal T}_3$ lattice. The analytical expressions of the pseudo-Landau levels (pLLs) are derived based on low-energy Hamiltonians, and are verified by tight-binding calculations. We find the pseudo-magnetic field leads to the oscillating density of states, and the first pLL is sublattice polarized, which is distributed on only two of the three sets of sublattices. For the nonuniform uniaxial strain, we show that pLLs become dispersive due to the renormalization of the Fermi velocity, and a valley polarized current emerges. Our results generalize the study of pseudo-magnetic field to the $\alpha-{\cal T}_3$ lattice, which will not only deepen the understanding of the intriguing effects of mechanical strains, but also provide theoretical foundation for possible experimental studies of the effects of strain on the $\alpha-{\cal T}_3$ lattice.
\end{abstract}


\maketitle

\section{Introduction}
The pseudo-magnetic field induced by strains in graphene has attracted extensive interests in recent years\cite{dejuan2013,vozmediano2010}. Its strength can be extremely large (up to $~800\textrm{T}$)\cite{hsu2020}, which is far beyond the current lab limit for a real magnetic field, thus providing probabilities to explore exotic quantum phenomena in giant magnetic fields. Various strain patterns have been proposed theoretically to generate pseudo-magnetic fields\cite{guinea2010a,  guinea2010b, neekamal2013, guinea2008, chang2012, zhang2014}, revealing many novel physical properties \cite{lantagne2020, oliva2020, lisy2020, settnes2016}. The existence of pseudo-magnetic fields has been repeatedly confirmed in  experiments by the scanning tunneling microscopy and the angle-resolved photoemission spectroscopy in mechanically deformed graphene\cite{levy2010, meng2013, lisy2015, hsu2020, nigge2019,jia2019}.  Beyond graphene, similar effects of pseudo-magnetic fields are also widely explored in other Dirac materials \cite{liu2017a, liu2017b, liu2019, liu2020, liu2021, filusch2021, cheng2021, naumis2017, sun2021a, sun2021b}.

\begin{figure}[t]
\centering
\includegraphics[width=8.6cm]{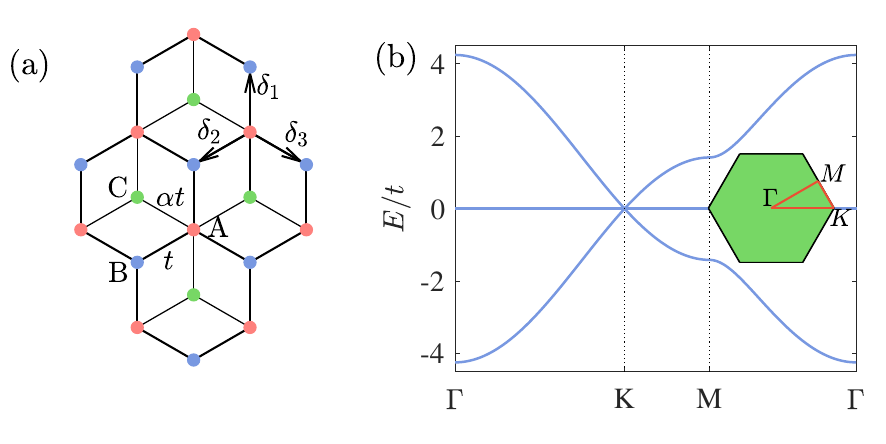}
\caption{(a) The schematic plot of the $\alpha-{\cal T}_3$ lattice. Each unit cell contains three sites labeled as $A, B, C$. $\bm \delta_i$ ($i = 1, 2, 3$) are the nearest-neighbor vectors. (b) Band structure of an infinite $\alpha-{\cal T}_3$ lattice with $\alpha=1$ along the high-symmetry path (i.e., the orange curve illustrated in the inset). Here the eigenenergy $E/t$ is measured in unit of $t$. Inset: the first Brillouin zone of the $\alpha-{\cal T}_3$ lattice with the high-symmetry points labeled.} \label{fig1}
\end{figure}

The $\alpha-{\cal T}_3$ lattice has a graphene-like geometry, which is formed by adding an extra site at the center of each hexagon of the honeycomb lattice, and connecting it to only one of the two inequivalent sublattices [Fig.~\ref{fig1}(a)] with a bond strength controlled by an anisotropic parameter $\alpha$ \cite{sutherland1986, raoux2014}. When $\alpha$ changes from $0$ to $1$, the $\alpha-{\cal T}_3$ lattice can continuously evolve from the honeycomb lattice ($\alpha=0$) to the dice lattice ($\alpha=1$) \cite{raoux2014, louvet2015, illes2015}. For $0<\alpha\leq 1$, the low-energy physics of the $\alpha-{\cal T}_3$ lattice is characterized by the same Dirac-Weyl Hamiltonian as graphene but with a larger pseudospin $S = 1$. The resulting band structure consists of two linear and one flat bands, intersecting at the Dirac points. Possible realizations of the $\alpha-{\cal T}_3$ lattice have been proposed in a variety of settings, ranging from electronic systems to artificial lattices \cite{okamoto2018, malcolm2015, bercioux2009, leykam2018, rizzi2006, abilio1999, naud2001, andrijauskas2015, wang2011}.

The $\alpha-{\cal T}_3$ lattice exhibits interesting physical properties (e.g., super-Klein tunneling\cite{illes2017, betancur2017}, novel topological phase\cite{wang2011, andrijauskas2015, dey2019, wang2021, dey2020}), among which the unconventional quantum Hall effect (QHE) is particularly intriguing \cite{bugaiko2019, xu2017, wang2020, illes2015}. With $\alpha$ increased, there is an evolution from the relativistic QHE to the nonrelavistic QHE. An unconventional QHE appears for $0<\alpha<1$, which can be regarded as a superposition of two extrema of $\alpha=0$ and $\alpha=1$. Here the Landau levels are not identical in the two valleys, and there exists an offset between them. The resulting Hall plateaus are quantized at $0,2,4,6,8,...$ in units of $e^2/h$, which is a combination of the relativistic quantization ($2,6,...$) and the nonrelavistic quantization ($0,4,8,...$). Such a behavior has been linked to the evolution of the Berry phase in the system, which changes continuously from $\pi$ ($\alpha=0$) to $0$ ($\alpha=1$) \cite{illes2015}. This peculiar feature (i.e., the offset) of the magnetic field can be used to realize the valley filtering in the Fabry-P\'erot interferences, making them promising for valleytronics applications \cite{xuhy2017, bouhadida2020}.

Considering the interesting effects of the strains in graphene and the similarity between the $\alpha-{\cal T}_3$ and honeycomb lattices, it is natural to ask what effects various nonuniform strain patterns may bring into the $\alpha-{\cal T}_3$ Hamiltonian. It is noted that except the strain effect from an out-of-plane Gaussian bump \cite{filusch2021}, the role of nonuniform strains is still largely unexplored in the $\alpha-{\cal T}_3$ lattice. In this manuscript, we investigate the pseudo-magnetic field in the $\alpha-{\cal T}_3$ lattice induced by two typical kinds of inhomogeneous strains, i.e., a non-uniform uniaxial strain and a triaxial strain, both of which have been known to generate uniform pseudo-magnetic fields in graphene. Based on the low-energy Hamiltonian and the tight-binding model, we calculate the strain-induced pLLs, and find the results from both approaches are in good consistence. We compare the pLLs and the Landau levels by a real magnetic field, and discuss their differences. A notable point is that the pLLs are dispersive, resulting in a valley polarized current through explicit calculations. Besides, the first pLL is sublattice polarized - only distributing on sublattices $A$ and $C$ of the $\alpha-{\cal T}_3$ lattice. Our results generalize the studies of pseudo-magnetic fields to the $\alpha-{\cal T}_3$ lattice, and thus will deepen the understanding of the intriguing effects of strains.

This manuscript is organized as follows. Section~\ref{sec2} introduces the strained Hamiltonian on the $\alpha-{\cal T}_3$ lattice. Section~\ref{sec3} studies the pLLs under an inhomogeneous uniaxial strain. Section~\ref{sec4} calculates the associated valley current in the uniaxially strained $\alpha-{\cal T}_3$ lattice. Section~\ref{sec5} presents the results of the pLLs induced by a triaxial strain. Finally, we enclose some further discussions and conclusions in Sec.~\ref{sec6}.

\section{The strained Hamiltonian}
\label{sec2}
We start from the following tight-binding Hamiltonian on the $\alpha-{\cal T}_3$ lattice, which is given by
\begin{equation}\label{H}
  H=-\sum_{n=1}^{3}t_n\left(\sum_{i\in B} c_i^\dagger c_{i-\delta_n}+\alpha\sum_{i\in C} c_i^\dagger c_{i+\delta_n}\right) + \rm H.c.,
\end{equation}
where $c_{i}^{\dagger}$ and $c_{i}$ are the creation and annihilation operators, respectively, at site $i$; $\alpha$ is the anisotropic parameter of the hopping amplitude; and $t_n=t$ is the hopping amplitude in the absence of strain along the bond $\bm\delta_n$. In the $\alpha-{\cal T}_3$ model, as schematically shown in Fig.~\ref{fig1}(a), $A$ and $B$ sublattices constitute a honeycomb lattice with hopping amplitude $-t$, and the $C$ sublattice located in the center of each hexagon is only connected to $A$ sublattice with amplitude $-\alpha t$. The parameter $\alpha$ can be continuously tuned from $0$ to $1$, evolving the lattice from a honeycomb lattice to a dice lattice.

After a Fourier transform, the Hamiltonian [Eq.~(\ref{H})] can be written as $H=\sum_{\bm{k}}\psi_{\bm{k}}^{\dagger} \mathcal{H}_{\bm{k}} \psi_{\bm{k}}$ with
\begin{equation}\label{Hk}
\mathcal H_{\bm k}=
\begin{pmatrix}
0 & \cos\varphi f_\bk & \sin\varphi f_\bk^*
\\
\cos\varphi f_\bk^* & 0 & 0
\\
\sin\varphi f_\bk & 0 & 0
\end{pmatrix},
\end{equation}
the basis $\psi_\bk=(c_{A,\bk},c_{B,\bk},c_{C,\bk})^\intercal$, $\tan\varphi=\alpha$, and
\begin{equation}
f_\bk=-\sum_{n=1}^3t_n e^{i\bk\cdot\boldsymbol{\delta}_n}.
\end{equation}
Here  the Hamiltonian has been rescaled by $\cos \varphi$ for convenience.
The whole spectrum comprises of one flat band $\varepsilon_{0}(\bk) = 0$, and two dispersive bands $\varepsilon_{\pm}(\bk) = \pm|f_\bk|$. The associated wave functions are
\begin{equation}\label{wf_flat}
  \left|\Psi_{0}\right\rangle=\left(\begin{array}{c}
0 \\
\sin \varphi e^{-i \theta_{\bk}} \\
-\cos \varphi e^{i \theta_{\bk}}
\end{array}\right)
\end{equation}
for the flat band and
\begin{equation}
  \left|\Psi_{\pm}\right\rangle=\frac{1}{\sqrt{2}}\left(\begin{array}{c}
\pm1 \\
\cos \varphi e^{-i \theta_{\bk}} \\
\sin \varphi e^{i \theta_{\bk}}
\end{array}\right)
\end{equation}
for the dispersive conduction and valence bands, where $\theta_\bk=\arg f_\bk$ is the angle of $f_\bk$. The dispersions near the corners of the Brillouin zone (BZ) are $\varepsilon_{\pm}(\bk) =\pm\hbar v_F k$, where the Fermi velocity reads $v_F=3ta/2\hbar$ and the lattice constant is set to $a=1$, constituting Dirac cones. Unlike the energy spectrum of graphene, there is an additional flat band intersecting the Dirac points. Thus the low-energy spectrum is characterized by a quasi-relativistic equation for spin-$1$ fermions.

\begin{figure}[t]
\centering
\includegraphics[width=8.6cm]{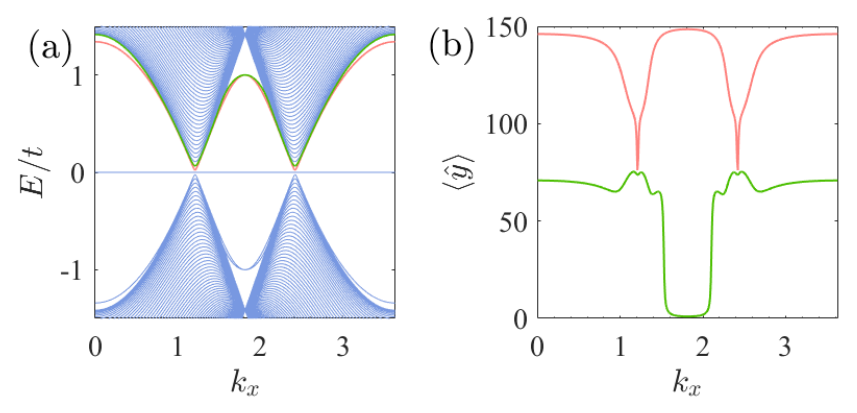}
\caption{(a) Band structure of an $\alpha-{\cal T}_3$ lattice nanoribbon in the absence of strain. (b) The expectation values of the position operator $\hat{y}$ for the peeled bands [red and green curves in panel (a)] demonstrate their edge-state nature. Here the system size is $L_y=100$, and the anisotropic parameter is $\alpha=1$.}\label{fig2}
\end{figure}

Figure~\ref{fig2}(a) plots the band structure of an undeformed $\alpha-{\cal T}_3$ lattice nanoribbon with $A-B$ zigzag boundaries. There appear four bands [colored lines in Fig.~\ref{fig2}(a) mark the upper two bands] peeled off from the bulk, and they are symmetric about the flat band at the Fermi energy. To reveal their properties, we calculate the average value of the $\hat{y}$ position operator for each state, $\langle \hat{y}\rangle=\langle \psi^{\dagger}_{\bf k}|\hat{y}|\psi_{\bf k}\rangle$, where $\psi_{\bf k}$ is the corresponding wave function at the momentum ${\bf k}$. As shown in Fig.~\ref{fig2}(b), the green band between the Dirac points is mainly distributed near the lower boundary, implying its edge-state nature. In contrast, $\langle\hat{y} \rangle$ of the red band takes the coordinate value of the upper boundary when $k_x$ is away from the Dirac points, thus corresponding to an upper-edge state. The two symmetric bands below the flat band are also edge states, whose distributions are the same with their electron-like counterparts.

The flat band is composed of a multitude of degenerate states, and the degeneracy is equal to the difference between the total number of $B,C$ sites and the number of $A$ sites in one unit cell, i.e., $N_d=N_B+N_C-N_A$, which is nicely consistent with Lieb’s theorem\cite{lieb1989}. Besides, the states in the flat band only distribute on the $B$ and $C$ sublattices, which is evidenced by the wave function of the flat band [Eq.~(\ref{wf_flat})]. In the real space, such states can be constructed locally within the $B,C$ sites of each unit cell, on which the wave function has equal amplitudes and opposite signs.

We then study the effect of an inhomogeneous uniaxial strain along the $y$ direction in a nanoribbon of an $\alpha-{\cal T}_3$ lattice. After the inhomogeneous uniaxial strain is applied, as a result of the lattice distortion, the hopping amplitudes will be modified to
\begin{equation}
t_n = t+\delta t_n,
\end{equation}
where $\delta t_n=t_{\bR^{\prime}+\bU(\bR^\prime),\bR+\bU(\bR)}-t_{\bR^\prime,\bR}$ is the hopping energy variation in the presence of strain deformation with $t_{\bR^\prime,\bR}=t$ being the hopping amplitude between sites located at $\bR$ and $\bR^\prime=\bR+\bm\delta_n$ and $\bU(\bR)$ being the displacement of the site at $\bR$. The hopping energy variation $\delta t_{n}$ can be recasted in terms of the strain tensor\cite{castroneto2009} $u_{ij}=(\partial_j U_i+\partial_i U_j)/2$ as
\begin{equation}\label{dt}
\begin{split}
\delta t_n&\approx \left[\bU(\bR^\prime)-\bU(\bR)\right]\cdot \bm \nabla_{\br^\prime-\br}t_{\br^\prime,\br}|_{\bR^\prime-\bR} \nonumber
\\
&\approx (\bR^\prime-\bR)\cdot\nabla_\br \bU(\br)|_\bR \cdot \frac{(\bR^\prime-\bR)}{|\bR^\prime-\bR|^2} (-\beta) t \nonumber
\\
&=-\frac{\beta t}{a^2}\bm{\delta}_n\cdot \bm{u}\cdot\bm{\delta}_n,
\end{split}
\end{equation}
where $\beta=-\partial \ln t_{\br^\prime,\br}/\partial \ln|\br^\prime-\br||_{|\br^\prime-\br|=a_0}$ is referred to as the Gr\"uneisen parameter\cite{vozmediano2010}.

\section{Unianxial strain-induced pseudo-Landau levels}
\label{sec3}
{
We present the analytical dispersions of the pLLs induced by a uniaxial strain characterized by $\bm U=(0,\tfrac{c}{2\beta}y^2)$. It is straightforward to check that the only nonvanishing component of the strain tensor is $u_{yy}=\tfrac{c}{\beta}y$, which leads to the hopping amplitudes
\begin{equation} \label{tn_uni}
\begin{split}
t_1&=t \left(1-cy \right),
\\
t_2&=t \left(1-\frac 1 4 cy \right),
\\
t_3&=t \left(1-\frac 1 4 cy \right).
\end{split}
\end{equation}
The effective Hamiltonian of the uniaxially strained $\alpha-{\cal T}_3$ lattice can then be obtained by incorporating Eq.~(\ref{tn_uni}) into Eq.~(\ref{H}). The resulting Bloch Hamiltonian reads $\mathcal{H}_{\bk}=\boldsymbol{h}(\boldsymbol{k})\cdot \boldsymbol{S}$, where $\bm h=(h_x,h_y)$ and $\bm{S}=\left(S_x,S_y\right)$ respectively are
\begin{equation}
\begin{split}
S_x&=
\begin{pmatrix}
0 & \cos\varphi & \sin\varphi
\\
\cos\varphi & 0 & 0
\\
\sin\varphi & 0 & 0
\end{pmatrix},
\\
S_y&=
\begin{pmatrix}
 0 & -i\cos{\varphi} & i\sin\varphi
\\
i\cos{\varphi} & 0 & 0
\\
-i\sin\varphi & 0 & 0
\end{pmatrix},
\end{split}
\end{equation}
and
\begin{equation}
\begin{split}
&h_{x}(\boldsymbol{k})=-\left(t_{1} \cos k_{y}+2 t_{2} \cos \frac{\sqrt{3} k_{x}}{2} \cos \frac{k_{y}}{2}\right), \\
&h_{y}(\boldsymbol{k})=-\left(-t_{1} \sin k_{y}+2 t_{2} \cos \frac{\sqrt{3} k_{x}}{2} \sin \frac{k_{y}}{2}\right).
\end{split}
\end{equation}
Writing the momentum near the Dirac point $\boldsymbol{K}^\pm=\left(\pm \frac{4\pi}{3\sqrt 3},0\right)$ as $\boldsymbol{k}= \bK^{\pm}+\boldsymbol{q}$, and expanding $h_{x}(\boldsymbol{k}), h_{y}(\boldsymbol{k})$ to linear order of $q_x,q_y$, the resulting Hamiltonian is,
\begin{equation} \label{hq_uni}
\begin{split}
h^\pm_{x}(\boldsymbol{q})& =\hbar v_F\left[\pm\left(1-\frac{c}{4}y\right)q_x \pm \frac{c}2y\right],
\\
h^\pm_{y}(\boldsymbol{q}) &=\hbar v_F\left[\left(1-\frac{3c}{4}y\right)q_y\pm\frac{q_xq_y}{2}\left(1-\frac{c}4y\right)\right],
\end{split}
\end{equation}
where the superscript ``$\pm$'' on the left-hand side refers to the two valleys $\boldsymbol{K}^\pm$. From the above low-energy effective Hamiltonian, it is clear that the uniaxial strain produces a pseudo-magnetic field with the vector potential $(\mathcal{A}_x,\mathcal{A}_y)=(\pm cy/2,0)$, which takes opposite signs at the two Dirac points and thus differs from the real magnetic field. In addition, the Fermi velocities along the $q_x$ and $q_y$ directions are renormalized, and thus become inhomogeneous over the lattice.
}

We now study the low-energy effective effective Hamiltonian
\begin{equation} \label{h_op}
\mathcal{H}^\xi(\bm{q})=\hbar v_F\left[(q+pcy)S_x+(s-rcy)(-i\partial_y)S_y\right],
\end{equation}
which is obtained by making a substitution $q_y\rightarrow -i\partial_y$ to Eq.~(\ref{hq_uni}). For transparency, we have define $p=\tfrac{1}{2}-\tfrac{1}{4}q$, $s=1+\tfrac 1 2 q$, and $r=\tfrac{3}{4}+\tfrac{1}{8}q$, where $q=\xi q_x$ with $\xi=\pm$ specifying the valleys. The eigenvalue problem of the effective Hamiltonian $\mathcal H^\xi(\bm q)$ can be mapped to a second-order ordinary differential equation (see Appendix~\ref{a1}), and the spectrum is given by
\begin{equation} \label{pLL_uni}
E^\xi_n=\pm\hbar v_F\sqrt{(n-\frac12-\text{sgn}(c)\frac{\cos2\varphi}{2})|c|\frac{2+3\xi q_x}{2}},
\end{equation}
where $\xi=\pm$ labels the two valleys; the pseudo-Landau level index $n\geq 1$; and $\cos2\varphi=\frac{1-\alpha^2}{1+\alpha^2}$.

From the above analytical expressions, it is found that the pLLs are dispersive with $q_x$. Since the spectra are only different in the signs in front of $q_x$ at the two valleys, they are mirror-symmetric about the $y$ axis, and the Fermi velocities are opposite in the vicinity of $\bm{K}^\pm$. When $\alpha$ is tuned between $0$ and $1$, the positions of pLLs change accordingly. The changing directions are opposite when the pseudo-magnetic field is reversed, implying the pLLs produced by tensile and compressive strains are different. It should be noted that the $\alpha=0$ limit corresponds to the normal honeycomb lattice, where $\cos2\varphi=1$ and the $\sqrt{n}$ dependence of the pLLs is recovered. In the other limit $\alpha=1$, the $\alpha-{\cal T}_3$ lattice is reduced to a Dice lattice. We have $\cos2\varphi=0$, and the positions of the pLLs are proportional to the square roots of half-integers.

\begin{figure}[t]
\centering
\includegraphics[width=8.6cm]{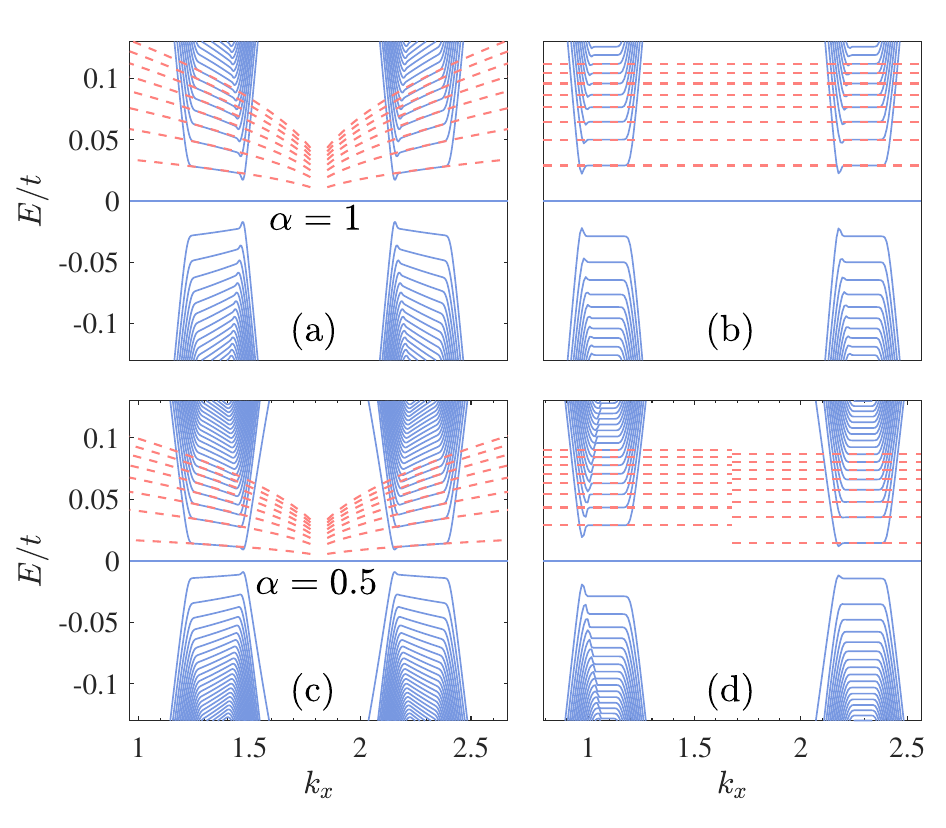}
\caption{The pseudo-Landau levels of the $\alpha-{\cal T}_3$ lattice nanoribbon under an inhomogeneous uniaxial strain for (a) $\alpha=1$ and (c) $\alpha=0.5$. (b) and (d) are the Landau levels induced by real magnetic fields with the same magnitudes as those in (a) and (c). The blue solid curves in (a)-(d) are obtained by diagonalizing the tight-binding model Hamiltonian of a uniaxially strained $\alpha-{\cal T}_3$ lattice nanoribbon with $L_y=900$, and the red dashed curves are the analytical dispersions obtained from the low-energy effective Hamiltonians. The pseudo-magnetic field (real magnetic field) corresponds to a vector potential $(\mathcal{A}_x,\mathcal{A}_y)=(\pm cy/2,0)$ [$(\mathcal{A}_x,\mathcal{A}_y)=(cy/2,0)$]. The parameter $c=0.5c_{max}$, where $c_{max}$ corresponds to the maximum strain which can be applied to the lattice.}\label{fig3}
\end{figure}

The above analytic results can be verified by diagonalizing the strained tight-binding Hamiltonian numerically, in which the changes of the hopping amplitudes along the three basic directions according to Eq.~(\ref{dt}) are: $\delta t_1=-cyt, \delta t_2=\delta t_3=-cyt/4$, respectively. For a fixed size $L_y$, there exists a maximum strain strength $c_{max}$, which is defined to ensure all $t_n>0$ [Eq.~(\ref{tn_uni})]. Hence we have $c_{max}=1/y_{max}$ with $y_{max}=\frac12+\frac32(L_y-1)$ being the coordinate of the $A$ sites on the upper boundary. It is worth noting that the details of the boundaries are irrelevant since the pLLs correspond to extended states in the bulk. The following calculations are performed on the $\alpha-{\cal T}_3$ nanoribbons terminated by $A-B$ zigzag boundaries.

As illustrated in Figs.~\ref{fig3}(a) and~\ref{fig3}(c), the analytical results and tight-binding calculations agree very well with each other in the vicinity of Dirac points. As a comparison, we also present the results of a real magnetic field $\bB=\tfrac{c}{2}\hat z$ with the same magnitude as that of the pseudo-magnetic field [Figs.~\ref{fig3}(b) and~\ref{fig3}(d)]. For the $\alpha=1$ case, the Landau levels generated by a pseudo-magnetic field and a real magnetic field are similar, except that the real Landau levels are dispersionless in the vicinity of the Dirac points. In contrast, when the anisotropic parameter $\alpha$ deviates from $1$ [for example $\alpha=0.5$ in Fig.~\ref{fig3}(d)], the positions of real Landau levels becomes different in two valleys. The reason lies in the analytical formula of the Landau levels: $E_n^{\xi}=\pm \hbar\upsilon_{F}\sqrt{(n-\frac{1}{2}-\text{sgn}(c)\xi\frac{\cos 2\varphi}{2})|c|}$, where the difference results from the additional sign [cf., Eq.~(\ref{pLL_uni})] in front of the term $\frac{\cos 2\varphi}{2}$ for the two valleys.

\begin{figure}[t]
\centering
\includegraphics[width=8.6cm]{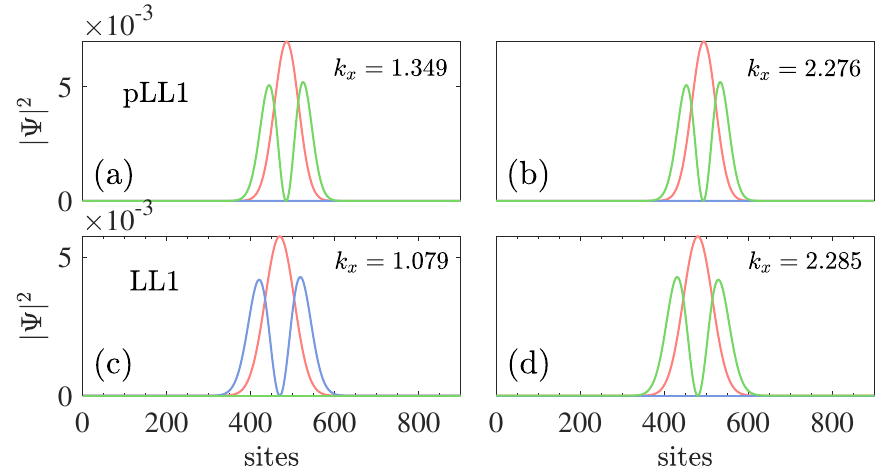}
\caption{The distribution of the wave function of the first pLL at momenta near the Dirac points: (a) $k_x=1.349$, (b) $k_x=2.276$. Similar plots under a real magnetic field at momenta: (c) $k_x=1.079$, (d) $k_x=2.285$. The red, blue, and green solid curves represent the distributions on the three sets of sublattices $A$, $B$, and $C$, respectively. Here the anisotropic parameter is $\alpha=1$, and the magnetic field strength is $B=c/2$ with $c=0.5c_{max}$.}\label{fig4}
\end{figure}

It is well known that the zeroth pLL is sublattice polarized in graphene \cite{shi2021, liu2022}. It is in great contrast to its counterpart in the real magnetic field, which is supported by both sublattices. Here we find a similar sublattice polarization for the first pLL in the strained $\alpha-{\cal T}_3$ lattice. As shown in Figs.~\ref{fig4}(a) and~\ref{fig4}(b), the first pLL in both valleys only distributes on the $A$ and $C$ sublattices. Here the plots are for the Dice lattice $(\alpha=1)$, but the phenomenon of the sublattice polarization remains valid for other values of $\alpha$. We also plot the distribution of the first Landau level induced by a real magnetic field in Figs.~\ref{fig4}(c) and~\ref{fig4}(d), which sits on the $A$ and $B(C)$ sublattices for the left (right) valley. Therefore, the first Landau level is distributed on all three sets of sublattices \cite{bugaiko2019, illes2016}. Unlike the first real Landau level, the first pLL only resides on two sets of sublattices (i.e., $A$ and $C$), exhibiting sublattice polarization. Finally, it is worth to emphasize the $\alpha=0$ case, where the $C$ sublattice becomes completely isolated from the lattice, and the situation in graphene is exactly recovered.

\section{Valley currents in the uniaxially strained \texorpdfstring{$\alpha-{\cal T}_3$}{} lattice}
\label{sec4}
In the previous section, we have proved that the inhomogeneous uniaxial strain can produce dispersive pLLs. Since the electrons near the two valleys have non-zero opposite group velocities, it is expected that a valley current will be generated in the strained $\alpha-{\cal T}_3$ lattice. For ballistic transport, the valley current in equilibrium is defined as
\begin{equation}
  I^{v}(\mu)=\sum_{\xi=\pm} \sum_{n} \int \frac{d k_{x}}{2 \pi}\left[\xi f\left(E_{n}^{\xi}\left(k_{x}\right)\right) v_{n}^{\xi}\left(k_{x}\right)\right],
\end{equation}
where $\xi=\pm$ labels the two valleys; $v_{n}^{\xi}\left(k_{x}\right)$ represents the group velocity of electrons on the $n$th band, which is,
\begin{equation}
  v_{n}^{\pm}\left(k_{x}\right) \equiv \frac{1}{\hbar} \frac{\partial E_{n}^{\pm}}{\partial k_{x}},
\end{equation}
and $f\left(E_{n}^{\xi}\right)=1/(e^{(E_n^\xi-\mu)/k_BT}+1)$ is the Fermi-Dirac distribution function at temperature $T$. In the following calculations, we takes the center of the Brillouin zone, i.e., $k_x=\pi/\sqrt{3}$, as the reference point, and divide the entire Brillouin zone into two parts, which correspond to the two different valleys. To resolve the real-space distribution of the valley current, the expectation value of the position operator $\langle\hat{y}\rangle$ for each eigenstate at the momentum $k_x$ is calculated. Based on the resulting expectation value, the contribution to the valley current is classified into three categories: bottom edge, top edge, and bulk.

\begin{figure}[t]
\centering
\includegraphics[width=8.6cm]{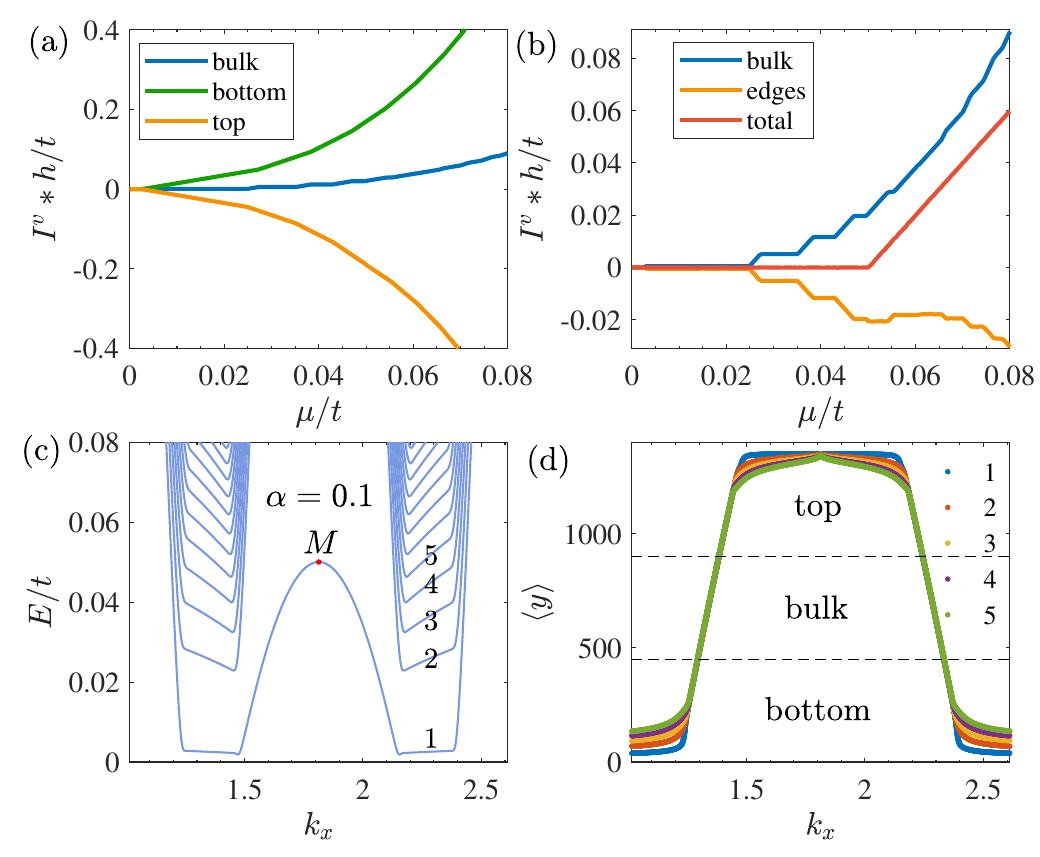}
\caption{(a) Equilibrium valley currents in three different parts of the nanoribbon as a function of $\mu$. (b) The bulk/edge contribution and the total valley current as a function of $\mu$. (c) Band structure of the strained $\alpha-{\cal T}_3$ lattice nanoribbon with the anisotropic parameter $\alpha=0.1$. (d) The expectation values of the position operator for the five lowest pLLs, according to which the eigenstates are classified into three categories: bottom-edge, top-edge, and bulk states. Here the system size is $L_y=900$ and the strain strength is $c=0.5c_{max}$.} \label{fig5}
\end{figure}

Figure~\ref{fig5}(a) plots the valley current in different parts of the nanoribbon as a function of $\mu$. The bulk contribution becomes non-zero at $\mu/t=0.0245$ and exhibits a series of plateaus afterwards. The bulk valley current increases linearly between adjacent plateaus. In contrast, the top and bottom contributions vary monotonically with $\mu$. Additionally, their signs are different, suggesting that the valley currents near the two edges move in opposite directions. The above behaviors can be well understood from the band structure shown in Fig.~\ref{fig5}(c). The bulk valley current is contributed by the pLLs. Since the first pLL is almost dispersionless, $I^{v}_{bulk}$ becomes non-zero only when the second pLL begins to be occupied, corresponding to the chemical potential $\mu/t=0.0245$. As $\mu$ increases, more states in this pLL become occupied, thus $I^{v}_{bulk}$ increases accordingly. When the second pLL is fully occupied, $I^{v}_{bulk}$ is intact, exhibiting a plateau, whose width is equal to the gap size between the second and the third pLLs. The states aside the pLLs are the edge states. Taking the right valley as an example, the states right (left) to the pLLs are mainly distributed near the bottom (top) edge [Fig.~\ref{fig5}(d)]. Since these states are continuously occupied with increasing $\mu$, the curves of $I^{v}_{edges}$ vary monotonically. The velocities of the bottom and top states have opposite signs; so do the corresponding valley currents.

As shown in Fig.~\ref{fig5}(b), the net valley current becomes nonzero only when $\mu/t$ is greater than the highest point [$M$ at $k_x=\pi/\sqrt{3}$ in Fig.~\ref{fig5}(c)] of the boundary state connecting the two segments of the first pLL. The emergence of the valley current arises from the imbalance of the co-propagating modes between the two valleys. For example, the right valley has an additional right-moving channel than the left valley when $\mu>E_M$, resulting in a net valley current.

\section{Triaxial strain-induced pseudo-Landau levels}
\label{sec5}
\begin{figure}[t]
\centering
\includegraphics[width=8.6cm]{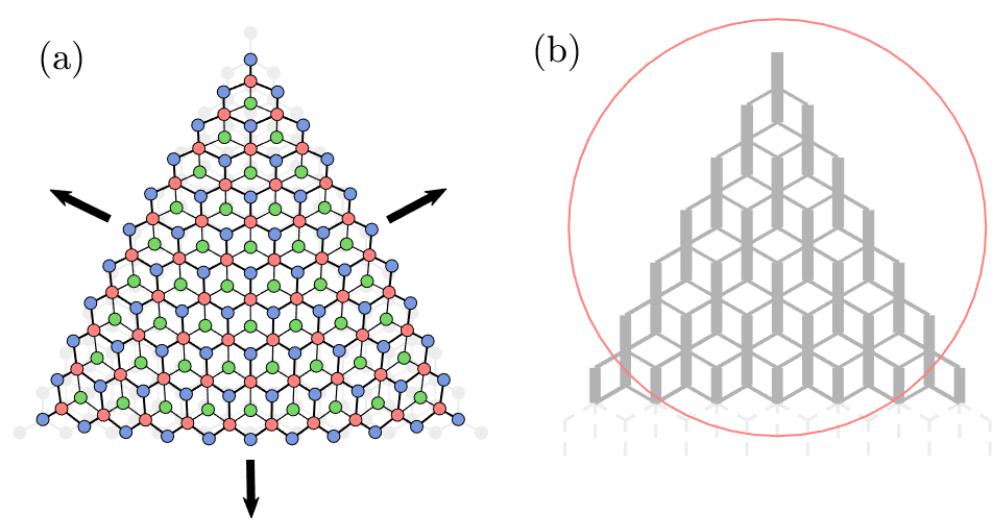}
\caption{(a) A schematic plot of a triaxially deformed $\alpha-{\cal T}_3$ lattice flake. The linear size is represented by the number of $B$ sites (blue solid circles) on each edge, which is $L=11$ for the present panel. The numbers of sites in sublattices $A$, $B$, and $C$ are unequal, and are $L(L-1)/2$, $L(L+1)/2$, and $(L-1)(L-2)/2$, respectively. (b) Enlargement in the vicinity of the upper corner of a strained triangular flake of size $L=50$. The value of the hopping on each bond is represented by the thickness of the bond. $c/c_{max} = 0.5$ is used as the strain strength in (b).} \label{fig6}
\end{figure}

We next consider a triaxial strain applied on the $\alpha-{\cal T}_3$ lattice, as schematically shown in Fig.~\ref{fig6}(a). The displacement function for a triaxial strain \cite{guinea2010a, neekamal2013} reads
\begin{equation}
  \bm{U}(x,y)=\frac{c}{\beta}(2xy, x^2-y^2).
\end{equation}
Then the hopping amplitudes become position-dependent as
\begin{equation}\label{tsub_tri}
\begin{split}
t_1=t+\delta t_1&=t(1+2cy),
\\
t_2=t+\delta t_2&=t\left[1-c(y+\sqrt3 x)\right],
\\
t_3=t+\delta t_3&=t\left[1-c(y-\sqrt3 x)\right].
\end{split}
\end{equation}
The bonds vertical to the edges are the most weakened, and thus determine a maximum strain strength, at which these bonds vanish. We adopt the triangular flake considered in Ref.\cite{poli2014} with all three edges being $A$-$B$ zigzag [Fig.~\ref{fig6}(a)]. The size of the flake is determined by the number of the $B$ sites on each edge, labeled as $L$. The maximum strain strength is then $c_{max}=1/(L-1)$.

First, we analytically solve the pLLs induced by a triaxial strain. Substituting the explicit forms of the hopping amplitudes [Eq.~(\ref{tsub_tri})] into the Hamiltonian [Eq.~(\ref{H})] and expanding to linear order of the momenta relative to $\bm{K}^\pm$, the effective Hamiltonian around the Dirac points under a triaxial strain can be obtained as $\mathcal{H}^\pm_{\bm{q}}=\bm h(\bm{q})^\pm\cdot \bm{S}$ with $\bm h=(h_x,h_y)$ and $\bm{S}=(S_x, S_y)$, where
\begin{equation}
\begin{split}
h_{x}^\pm(\boldsymbol{q})&=\hbar v_F\left[-2cy\pm(1-cy)q_x\pm(-cx q_y) \right],
\\
h_{y}^\pm(\boldsymbol{q})&=\hbar v_F\left[\pm2cx-cx q_x+(1+cy)q_y\right].
\end{split}
\end{equation}
The low-energy Hamiltonian can be rewritten as
\begin{align}\label{hq_tri}
\mathcal{H}^\pm(\bm{q})=\hbar \bm{S}\cdot\bar{\bm{v}}^\pm\cdot (\boldsymbol{q}+ \boldsymbol{\mathcal{A}}^\pm) ,
\end{align}
where $\bm{\mathcal{A}}^\pm=({\mathcal{A}}^\pm_x,{\mathcal{A}}^\pm_y)$ with
\begin{align*}
&{\mathcal{A}}^\pm_x=\pm \cdot \frac{2 c^2 x^2-2 c y(1+c y)}{1-c^2 x^2-c^2 y^2} \\
&{\mathcal{A}}^\pm_y=\pm \cdot \frac{2 c x(1-2 c y)}{1-c^2 x^2-c^2 y^2},
\end{align*}
and
\begin{align}
\boldsymbol{\bar{v}}^\pm=v_F\left(
                          \begin{array}{cc}
                            \pm(1-cy) & \pm(-cx) \\
                            -cx & 1+cy \\
                          \end{array}
                        \right).
\end{align}
From the above formulas, it is direct to see that the triaxial strain results in two effects: renormalizing the Fermi velocity\cite{dejuan2012} with a tensor $\bar{\bm{v}}^\pm$ and generating a vector potential $\bm{\mathcal A}^\pm$.
For small values of the strain strength $c$, the Fermi velocity may be regarded as constant, and the vector potential approximates to $\bm{\mathcal{A}}^\pm\approx\pm(-2cy,2cx)$, when the pseudo-Landau levels can be solved analytically.

Solving the eigenproblem of the Hamiltonian $\mathcal{H}^\pm(\bm q)$ (see Appendix~\ref{a2}), we obtain the pLLs generated by the triaxial strain,
\begin{equation}\label{pLL_tri}
E_n=\pm\hbar v_F\sqrt{\left(n-\frac12+\text{sgn}(c)\frac{\cos2\varphi}{2}\right)|8c|},\ n\geq1,
\end{equation}
where the pLLs are again mirror-symmetric about the $y$ axis [cf., Eq.~(\ref{pLL_uni})].

\begin{figure}[t]
\centering
\includegraphics[width=8.6cm]{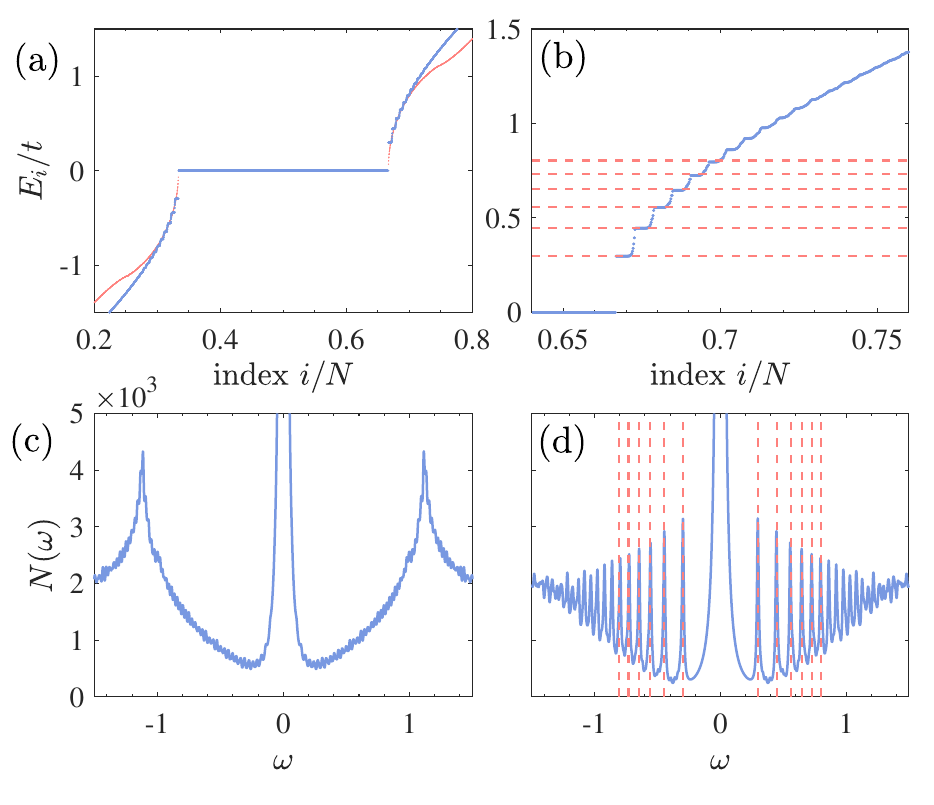}
\caption{Eigenvalues and density of states of strained and unstrained triangular flakes of the $\alpha-{\cal T}_3$ lattice. (a) Eigenvalues of the strained (blue) and unstrained (red) triangular flakes. (b) Enlargement of the pLLs for $E>0$ in panel (a). (c) and (d) are the corresponding densities of states of the unstrained and strained triangular flakes, respectively. The system size is $L=101$ and the anisotropic parameter is $\alpha=0.5$. The strain strength in panels (a), (b) and (d) is $c=0.5c_{max}$. The red dashed curves in panels (b) and (d) are from the analytical results in Eq.~(\ref{pLL_tri}).} \label{fig7}
\end{figure}

Then, we verify the above analytical results by directly diagonalizing the tight-binding Hamiltonian under the triaxial strain. Since the triaxial strain breaks the translational symmetry in both $x$ and $y$ directions, we can only obtain a set of discrete energy eigenvalues in real space. The red curve in Fig.~\ref{fig7}(a) shows the unstrained eigenenergies sorted by ascending order. A highly degenerate plateau at $E=0$ is apparent, which corresponds to the flat band [Fig.~\ref{fig1}(b)] of the $\alpha-{\cal T}_3$ lattice (here $\alpha=0.5$). The other eigenvalues come from the dispersive bands [Fig.~\ref{fig1}(b)].  After a triaxial strain is applied, there appear a series of small plateaus [blue dotted curve, Fig.~\ref{fig7}(a)] in the two $E\neq 0$ branches, which reflect the strain-induced pLLs. We plot the analytical positions of the pLLs in Fig.~\ref{fig7}(b), and find a good consistency between the analytics and numerics. To demonstrate the pLLs more clearly, the density of states is calculated as a function of energy $\omega$, which is defined as
\begin{equation}
  N(\omega)=\sum_i\delta(\omega-E_i).
\end{equation}
In practical calculations, the Lorentz function is used to approximate the $\delta$ function
\begin{equation}
  \delta(x)\approx \frac{1}{\pi}\frac{\eta}{x^2+\eta^2},
\end{equation}
where we have used a broadening of $\eta=0.008t$ in the calculations.

\begin{figure}[t]
\centering \includegraphics[width=8.6cm]{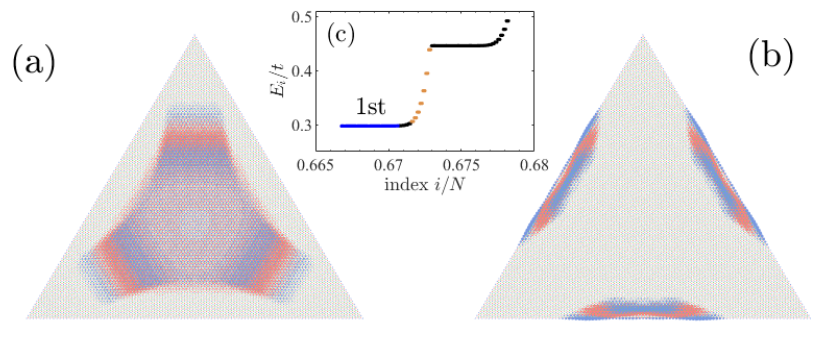} \caption{The distributions of the wave functions: (a) the states in the first pLL [blue dots in (c)]; (b) the states between the plateaus [yellow dots in (c)]. The colors in (a) and (b) represent the sublattices.}
\label{fig8}
\end{figure}

As shown in Fig.~\ref{fig7}(d), the pLLs are manifested in the density of states as sharp peaks, whose positions exactly match those of the pLLs. In contrast, we find $N(\omega)$ of an unstrained $\alpha-{\cal T}_3$ [Fig.~\ref{fig7}(c)] is in general smooth except at the flat band ($\omega=0$) and the Van Hove singularities ($\omega=\pm t\sqrt{1+\alpha^2}$). Thus the oscillating behavior of the density of states can be regarded as a direct evidence of the existence of the pLLs. To further reveal the properties of the pLLs, we calculate the real-space distributions of all the quantum states. Figure~\ref{fig8}(a) [\ref{fig8}(b)] shows the states on (between) the plateaus in Fig.~\ref{fig7}(b) are bulk (boundary) ones. Hence, in the energy spectrum [Fig.~\ref{fig7}(b)], the different bulk pLLs are connected by the dispersive boundary states.

\section{Conclusions}
\label{sec6}
Two types of nonuniform strains, which have been known to induce a homogenous pseudo-magnetic field in graphene, are investigated in the $\alpha-{\cal T}_3$ lattice. Specifically, we consider a linearly increasing uniaxial strain in a nanoribbon, and a triaxial strain in a triangular flake. For both cases, we derive the low-energy effective Hamiltonians under the strains, and obtain the analytical dispersions of the pLLs. It is found that the pLLs are proportional to $\sqrt{n+c_{\alpha}}$ with $c_{\alpha}$ being a constant depending on $\alpha$. Its connection to the $\sqrt{n}$ relation of graphene is obvious since $c_{\alpha}=0$ at the honeycomb point $\alpha=0$. The analytical expressions are further verified by directly diagonalizing the strained tight-binding Hamiltonians. For the uniaxially strained nanoribbon, we show that the first pLL is sublattice polarized and is only supported by the $A$ and $C$ sublattices of the $\alpha-{\cal T}_3$ lattice. Besides generating a vector potential, the uniaxial strain also renormalizes the Fermi velocity in the Hamiltonian, resulting in dispersive pLLs and a valley polarized current. It is worth noting that the $\alpha\neq 0,1$ case is unique in that the pLLs are always symmetric between the two valleys, which is in great contrast to the asymmetric Landau levels induced by a real magnetic field. For the triaxially strained flake, the existence of the pLLs is manifested by the induced oscillatory density of states.

Our study extends the strain-induced pseudo-magnetic field to the $\alpha-{\cal T}_3$ lattice, which will not only deepen the understanding of the underlying physics of pLLs, but also provide theoretical foundation for possible experimental realizations. Although there has been no report of two-dimensional quantum materials with perfect $\alpha-{\cal T}_3$ lattice yet, it is possible to create an artificial one in setups like Josephson junction arrays, photonic crystals, or electrical circuit networks \cite{polini2013, leykam2018, haenel2019, guzman2014, ningyuan2015}. Some of the above platforms can easily realize the desired strain patterns\cite{mann2020, rechtsman2013}, since the positions of the lattice sites therein can be displaced at need, and an external stretch/compression usually used to produce the strains is unnecessary.

\begin{acknowledgments}
The authors thank Wen Yang, Chenyue Wen, Xingchuan Zhu for helpful discussions. J.S and H.G. acknowledges support from the NSFC grant Nos.~11774019 and 12074022, the NSAF grant in NSFC with grant No. U1930402, the Fundamental Research
Funds for the Central Universities and the HPC resources
at Beihang University. Y.D. is financially supported by Beijing Natural Science Foundation (Z180007) and National Natural Science Foundation of China (12074021).
\end{acknowledgments}

\appendix
\section{Solution of pLLs under a nonuniform uniaxial strain}
\label{a1}
The effective Hamiltonian operator under a uniaxial strain is given by Eq.~(\ref{h_op}). For convenience, we do a variable substitution $y\rightarrow y+\frac{s}{rc}$, then Eq.~(\ref{h_op}) becomes
\begin{equation} \label{h_op2}
\mathcal{H}^\xi(\bm{q})=\hbar v_F\left\{\left[q+pc\left(y+\frac{s}{rc}\right)\right]S_x+{\rm i}rcy\partial_yS_y\right\}.
\end{equation}
Since ${\rm i}y\partial_y$ in Eq.~(\ref{h_op2}) is non-Hermitian, we replace it with a symmetric Hermitian operator ${\rm i}\frac{y\partial_y+\partial_y y}{2}={\rm i}(y\partial_y+\frac12)$. Assuming the wave function has the form $\Psi=e^{iq x}(\phi_A,\  \phi_B,\ \phi_C)^\intercal$, we obtain from the eigenproblem $\mathcal{H}^\xi({\bm q})\Psi=E\Psi$ the following equations
\begin{equation}
\begin{split}
&\cos\varphi\left\{\left[q+pc\left(y+\frac{s}{rc}\right)\right]+rc(y\partial_y+\frac12)\right\}\phi_B+
\\
&\sin\varphi\left\{\left[q+pc\left(y+\frac{s}{rc}\right)\right]-rc(y\partial_y+\frac12)\right\}\phi_C=\varepsilon\phi_A,
\\
&\cos\varphi\left\{\left[q+pc\left(y+\frac{s}{rc}\right)\right]-rc(y\partial_y+\frac12)\right\}\phi_A=\varepsilon\phi_B,
\\
&\sin\varphi\left\{\left[q+pc\left(y+\frac{s}{rc}\right)\right]+rc(y\partial_y+\frac12)\right\}\phi_A=\varepsilon\phi_C,
\end{split}
\end{equation}
where we have defined $\varepsilon=E/\hbar v_F$ for transparency. Eliminating $\phi_B$ and $\phi_C$, we arrive at a second-order differential equation for $\phi_A$,
\begin{equation} \label{ode}
\left(y^2\phi^{\prime\prime}_A+2y\phi^\prime_A\right)-\left[\frac{p^2}{r^2}y^2+\eta y+\frac{\Delta}{c^2r^4}-\frac{1}{4}\right]\phi_A=0,
\end{equation}
with
\begin{equation}
\begin{split}
\Delta&=(ps+qr)^2-r^2\varepsilon^2,
\\
\eta&=\frac{p}{cr^3}(2ps+2qr+\cos(2\varphi)cr^2).
\end{split}
\end{equation}
We first find the asymptotic solutions at $y\rightarrow 0,\pm \infty$. In the vicinity of $y=0$, we can neglect the two terms containing $y,y^2$ in the square bracket of Eq.~(\ref{ode}). Equation~(\ref{ode}) is then reduced to
\begin{equation}
\left(y^2\phi^{\prime\prime}_A+2y\phi^\prime_A\right)-\left(\frac{\Delta}{c^2r^4}-\frac{1}{4}\right)\phi_A=0.
\end{equation}
which is a Cauchy-Euler equation with a convergent solution $\phi_A\sim y^{-\frac{1}{2}+\frac{\sqrt{\Delta}}{|c|r^2}}$ in the $y\rightarrow0$ limit. As $y\rightarrow \pm\infty$, Eq.~(\ref{ode}) may be simplified as
\begin{equation}
\phi^{''}_A-\frac{p^2}{r^2}\phi_A=0,
\end{equation}
whose general solution $\phi_A=Ae^{-\frac{p}{r}y}+Be^{\frac{p}{r}y}$ cannot converge at $y \rightarrow -\infty$ and $y\rightarrow +\infty$ simultaneously. It is critically important to note that physical hopping amplitudes $t_n$ [Eq.~(\ref{tn_uni})] must be positive-definite. This requires $c\rightarrow 0^{\mp}$ [i.e., $\text{sgn}(c)=\mp$] when $y\rightarrow \pm\infty$. This implies that a convergent solution may be written as $\phi_A=e^{\text{sgn}(c)\frac p r y}$. With such an \textit{a posteriori} solution, we may write the full solution to Eq.~(\ref{ode}) as $\phi_A= e^{\text{sgn}(c)\frac p r y} y^{-\frac{1}{2} + \frac{\sqrt{\Delta}}{|c|r^2}} u(y)$, which, upon substituting to Eq.~(\ref{ode}), gives
\begin{equation} \label{ode2}
\begin{split}
&yu^{\prime\prime}(y)+\left[1+\frac{2\sqrt{\Delta}}{|c|r^2}+2\text{sgn}(c)\frac{p}{r}y\right] u^\prime(y)+ 2 \text{sgn}(c)\frac{p}{r}
\\
&\times \left[\frac{1-\text{sgn}(c)\cos(2\varphi)}{2}+\frac{(\sqrt{\Delta}-qr-ps)}{|c|r^2}\right]u(y)=0.
\end{split}
\end{equation}
For transparency, we define $\gamma=1+\frac{2\sqrt{\Delta}}{|c|r^2}$, $\alpha=\frac{1-\text{sgn}(c)\cos(2\varphi)}{2}+\frac{(\sqrt{\Delta}-qr-ps)}{|c|r^2}$, and $z=-2\text{sgn}\frac{p}{r}y$. Then Eq.~(\ref{ode2}) is reduced to
\begin{equation}
zu''(z)+(\gamma-z)u'(z)-\alpha u(z)=0
\end{equation}
which is a confluent hypergeometric equation with a regular singularity at $z=0$ and can be solved by the series expansion method. One solution is,
\begin{equation}
u(z)=1+\frac{\alpha}{\gamma}\frac{z}{1!}+\frac{\alpha(\alpha+1)}{\gamma(\gamma+1)}\frac{z^2}{2!}+\cdots, \gamma\neq 0,-1,-2,\cdots.
\end{equation}
To make $u(z)$ a polynomial (thus finite), $\alpha$ should be $0$ or negative integers, i.e., $\alpha=-n, n=0,1,2,\cdots$. Hence we obtain the following expression for the eigenenergy,
\begin{equation}
E^\xi_n=\pm\hbar v_F\sqrt{(n-\frac12-\text{sgn}(c)\frac{\cos2\varphi}{2})|c|\frac{2+3\xi q_x}{2}},\quad n\geq 1,
\end{equation}
where we have made a substitution $n\rightarrow n-1$, thus $n= 1,2,\cdots$; and $\cos2\varphi=\frac{1-\alpha^2}{1+\alpha^2}$.

\section{Solution of pLLs under a nonuniform triaxial strain}
\label{a2}
Here we solve the pLLs induced by a triaxial strain based on the effective Hamiltonian [Eq.~(\ref{hq_tri})]. For small strain strength $c$, we take the following approximations: $\boldsymbol{\mathcal{A}}^\pm\approx \pm(-2cy, 2cx)$ and $\boldsymbol{\bar{v}}^\pm\approx v_F\rm{diag}(\pm 1, 1)$. Since the triaxial strain breaks the translational symmetry in both $x$ and $y$ directions, $q_x$ and $q_y$ are no longer good quantum numbers. Making substitutions $q_x\rightarrow -{\rm i}\partial_x,\ q_y\rightarrow -{\rm i}\partial_y$, Eq.~(\ref{hq_tri}) becomes
\begin{align}
\nonumber
h_{x}^\pm(\boldsymbol{q})&=\hbar v_F\left[-2cy\pm(-{\rm i}\partial_x)\right], \\
h_{y}^\pm(\boldsymbol{q})&=\hbar v_F\left[\pm 2cx-{\rm i}\partial_y\right].
\end{align}
We then define $\Pi_x^\xi=-2cy+\xi(-{\rm i}\partial_x)$ and $\Pi_y^\xi=\xi 2cx-{\rm i}\partial_y$, whose commutation relation reads
 \begin{equation}\label{}
   [\Pi_x^\xi,\Pi_y^\xi]=-{\rm i}4c.
 \end{equation}
In terms of $\Pi_x^\xi$ and $\Pi_y^\xi$, the Hamiltonian $\mathcal{H}^\xi_{\bm{q}}$ can be rewritten as
\begin{widetext}
\begin{align}
\mathcal{H}^\xi_{\bm{q}}=\hbar v_F\left(
                                  \begin{array}{ccc}
                                    0 & \cos\varphi(\Pi_x^\xi-{\rm i}\Pi_y^\xi) & \sin\varphi(\Pi_x^\xi+{\rm i}\Pi_y^\xi) \\
                                    \cos\varphi(\Pi_x^\xi+{\rm i}\Pi_y^\xi) & 0 & 0 \\
                                    \sin\varphi(\Pi_x^\xi-{\rm i}\Pi_y^\xi) & 0 & 0 \\
                                  \end{array}
                                \right).
\end{align}
\end{widetext}
We have the following commutation relationship
 \begin{equation}\label{}
   [\Pi_x^\xi-{\rm i}\Pi_y^\xi,\Pi_x^\xi+{\rm i}\Pi_y^\xi]=8c,
 \end{equation}

In solving the eigenvalues of the Hamiltonian in Eq.(B3), it is convenient to define two bosonic operators,
 \begin{align}\label{}
 \hat{a}&=\frac{1}{\sqrt{|8c|}}\left(\Pi_x^\xi-{\rm i}\Pi_y^\xi\right),\\
 \hat{a}^\dagger&=\frac{1}{\sqrt{|8c|}}\left(\Pi_x^\xi+{\rm i}\Pi_y^\xi\right),
 \end{align}
which satisfy $[\hat{a},\hat{a}^\dagger]=1$ for $c>0$. So we obtain
\begin{align}
\mathcal{H}^\xi_{\bm{q}}=\hbar v_F\sqrt{8|c|}\left(
                                  \begin{array}{ccc}
                                    0 & \cos\varphi\hat{a} & \sin\varphi\hat{a}^\dagger \\
                                    \cos\varphi\hat{a}^\dagger & 0 & 0 \\
                                    \sin\varphi\hat{a} & 0 & 0 \\
                                  \end{array}
                                \right).
\end{align}
Assuming the eigenfunction of $\mathcal{H}^+_{\bm{q}}$ to be $\Psi=(\phi_A,\phi_B,\phi_C)^\intercal$,
and expanding the eigenproblem $\mathcal{H}^+_{\bm{q}}\Psi=E\Psi$, we obtain the following three equations
 \begin{align}\label{}
 \nonumber
&\cos\varphi\hat{a}\phi_B+\sin\varphi\hat{a}^\dagger\phi_C=\epsilon\phi_A,\\  \nonumber
&\cos\varphi\hat{a}^\dagger\phi_A=\epsilon\phi_B,\\
&\sin\varphi\hat{a}\phi_A=\epsilon\phi_C,
 \end{align}
where $\epsilon=E/(\hbar v_F\sqrt{|8c|})$. Eliminating $\phi_B$ and $\phi_C$, we get the relationship
 \begin{align}\label{}
\cos^2\varphi\hat{a}\hat{a}^\dagger\phi_A+\sin^2\varphi\hat{a}^\dagger\hat{a}\phi_A=\epsilon^2\phi_A.
 \end{align}
By a replacement $\hat{a}^\dagger\hat{a}=n$, we have $\epsilon=\pm\sqrt{n+\cos^2\varphi}$. Thus the analytical expression of the pLLs for $c>0$ is
 \begin{align}\label{}
E=\pm\hbar v_F\sqrt{\left(n-\frac12+\frac{\cos2\varphi}{2}\right)|8c|},\ n\geq1,
\end{align}
where $\cos2\varphi=\frac{1-\alpha^2}{1+\alpha^2}$. For $c<0$, by exchanging the definitions of $\hat{a}$ and $\hat{a}^\dagger$ in Eq.(B6-B7), it is direct to obtain
\begin{align}\label{}
E=\pm\hbar v_F\sqrt{\left(n-\frac12-\frac{\cos2\varphi}{2}\right)|8c|},\ n\geq1.
 \end{align}

%
%
%
%
%
%
%
\bibliography{diceref}

\end{document}